\documentclass[preprint]{iucr}[s,nohead] % DO NOT DELETE THIS LINE
\usepackage{amsmath}
\usepackage{soul,color}
\usepackage{hyperref}
\usepackage{lineno}
\usepackage{cite}
\usepackage[symbol]{footmisc}
\usepackage{setspace}
     \journalcode{S}              % Indicate the journal to which submitted
\begin{document}                  % DO NOT DELETE THIS LINE

     %-------------------------------------------------------------------------
     % The introductory (header) part of the paper
     %-------------------------------------------------------------------------

     % The title of the paper. Use \shorttitle to indicate an abbreviated title
     % for use in running heads (you will need to uncomment it).

\title{MARTApp: software for the processing and reconstruction of synchrotron radiation-based magnetic tomographies}
\shorttitle{MARTApp}
\onehalfspacing
     % Authors' names and addresses. Use \cauthor for the main (contact) author.
     % Use \author for all other authors. Use \aff for authors' affiliations.
     % Use lower-case letters in square brackets to link authors to their
     % affiliations; if there is only one affiliation address, remove the [a].

\cauthor[a,b,c]{A. Estela}{Herguedas-Alonso}{herguedasalicia@uniovi.es}{address if different from \aff}
\author[a]{Joaqu\'{i}n}{G\'{o}mez S\'{a}nchez}
\author[a]{Claudia}{Fern\'{a}ndez-Gonz\'{a}lez}
\author[a]{Andrea}{Sorrentino}
\author[a]{Salvador}{Ferrer}
\cauthor[a]{Eva}{Pereiro}{epereiro@cells.es}{address if different from \aff}
\cauthor[b,c]{Aurelio}{Hierro-Rodriguez}{hierroaurelio@uniovi.es}{address if different from \aff}

\aff[a]{ALBA Synchrotron Light Source, CELLS, Cerdanyola del Valles, 08290, Barcelona, \country{Spain}}
\aff[b]{Departamento de Física, Universidad de Oviedo, 33007, Oviedo, \country{Spain}}
\aff[c]{CINN (CSIC – Universidad de Oviedo), 33940, El Entrego, \country{Spain}}

     % Use \shortauthor to indicate an abbreviated author list for use in
     % running heads (you will need to uncomment it).

\shortauthor{Herguedas-Alonso, A.E. and G\'{o}mez S\'{a}nchez, J. and Fern\'{a}ndez-Gonz\'{a}lez, C. and Sorrentino, A. and Ferrer, S. and Pereiro, E. and Hierro-Rodriguez, A.}

     % Use \vita if required to give biographical details (for authors of
     % invited review papers only). Uncomment it.

%\vita{Author's biography}

     % Keywords (required for Journal of Synchrotron Radiation only)
     % Use the \keyword macro for each word or phrase, e.g. 
     % \keyword{X-ray diffraction}\keyword{muscle}


     % PDB and NDB reference codes for structures referenced in the article and
     % deposited with the Protein Data Bank and Nucleic Acids Database (Acta
     % Crystallographica Section D). Repeat for each separate structure e.g
     % \PDBref[dethiobiotin synthetase]{1byi} \NDBref[d(G$_4$CGC$_4$)]{ad0002}

%\PDBref[optional name]{refcode}
%\NDBref[optional name]{refcode}

\maketitle                        % DO NOT DELETE THIS LINE

%\linenumbers

%\begin{synopsis}
%Supply a synopsis of the paper for inclusion in the Table of Contents.
%\end{synopsis}

\begin{abstract}
Magnetic vector tomography allows for visualizing the 3D magnetization vector of magnetic nanostructures and multilayers with nanometric resolution. In this work, we present MARTApp (Magnetic Analysis and Reconstruction of Tomographies Application), a software designed to analyze the images obtained from a full-field or scanning transmission X-ray microscope and reconstruct the 3D magnetization of the sample. Here, its workflow and main features are described. Moreover, a synthetic test sample consisting of a hopfion is used to exemplify the workflow from raw images to the final 3D magnetization reconstruction.
\end{abstract}

     %-------------------------------------------------------------------------
     % The main body of the paper
     %-------------------------------------------------------------------------
     % Now enter the text of the document in multiple \section's, \subsection's
     % and \subsubsection's as required.

\section{Introduction}

In recent years, magnetic vector tomography (MVT) has emerged as a powerful imaging technique to reveal the 3D magnetization from 2D images or projections of a sample under different orientations \cite{phatak2010,donnelly2017nanotomo,hierro2020thinfilms,dipietro2023}. This technique has gained popularity due to its ability to visualize complex 3D magnetic configurations at the nanoscale, which could serve as information carriers in magnetic sensing and data storage devices \cite{pacheco2017mag3D,dieny2020spintronic,wolf2022}.

MVT can be performed using neutrons, electrons, or X-rays. With neutrons, it is possible to image the magnetic induction field for samples of several centimetres due to its penetration depth \cite{manke2010neutrons}. However, this technique is limited to resolutions on the order of microns \cite{kardjilov2008neutrons}. Electrons are also suitable to probe the 3D magnetic induction on thin samples with high spatial resolution, typically less than 100 nm due to the penetration depth of electrons \cite{phatak2015ltem}. Using $10^2-10^3$ nm X-rays (either soft or hard respectively) it is possible to detect the magnetization of samples with thicknesses up to 2 $\mu$m with nanometer resolution depending on the technique used \cite{neethirajan2024}. A key advantage of X-rays over neutrons or electrons lies in directly probing the magnetization with element sensitivity by exploiting the interaction between the magnetization and circularly polarized photons \cite{chen1990xmcd, fischer1997xmcd}. Moreover, transmission X-ray microscopy, as a photon-in/photon-out technique, is unsensitive to external magnetic fields, allowing for a great variety of sample environments \cite{sorrentino2015mistral}. For instance, pump-probe experiments using a 500 MHz RF magnetic field combined with laminography was used to study the magnetization dynamics in a GdCo microdisk revealing the movement of the vortex domain wall \cite{donnelly2020time}.
The implementation of MVT has been used to reconstruct and analyze 3D magnetic textures. For example, Bloch points were identified in Py microstructures and the emergent field connecting the topological charges was computed \cite{hermosa2023boomerang}. Moreover, using hard X-ray nanotomography, the magnetization of cylinders of GdCo$_2$ showing vorticity loops containing vortex-antivortex pairs or Bloch points were imaged \cite{donnelly2021vortex}.

Despite the MVT uniqueness as an experimental tool to observe the 3D magnetization vector, the analysis of the high amount of data acquired in order to perform the reconstruction is a major drawback for the unexperienced user community. Note that to obtain the magnetization vector field in the axial tomography configuration, it is necessary to acquire two orthogonal tilt series (TS) to access the three components of the magnetization. In addition, each one of these TS must be recorded with circular right ($C+$) and left ($C-$) polarizations in order to obtain the magnetic signal through the X-ray Magnetic Circular Dichroism signal (XMCD) \cite{stohr2006magnetism}. Moreover, to increase the signal-to-noise ratio, multiple projections per angle are often recorded under the same polarization to perform their average. Overall, a typical MVT experiment implies the acquisition of thousands of images (considering several repetitions per projection, two TS, and two polarizations; this for both sample and flat field images of the incoming beam) \cite{hierro2020thinfilms}. Altogether, the data processing needs to deal with a large number of images and the use of accurate alignment methods prior to vector field reconstruction algorithms. 

To the best of our knowledge, there are no open-source tool available for MVT that covers from the acquisition to the magnetic reconstruction although several graphical user interfaces (GUI) have been implemented to process images from TXM \cite{liu2012txmwizard, welborn2024tomopyui,xiao2022txmsandbox}. In order to address this absence of applications and make this technique approachable to non-expert users, we introduce MARTApp, a suite of pipelines accompanied by an intuitive GUI that processes recorded projections and enables the recovery of the magnetic signal in 3D. The software has been particularly designed to process images acquired using the MISTRAL transmission X-ray microscope (TXM) at the ALBA Synchrotron \cite{sorrentino2015mistral} but it can be extended to any beamline where a transmission scheme applies (i.e. Scanning-TXM\footnote[1]{Although in a STXM there are geometric limitations to perform magnetic tomography, the magnetic signal is obtained similarly to the TXM scheme}\footnote[2]{Also includes ptychography reconstructed transmission images \cite{donnelly2016ptycho}} and TXM).
In the following sections, we describe the workflow of MVT in a TXM, the software design, the use cases providing a concrete example, and finally the features that MARTApp provides to the user community.

\section{Scientific background}

\begin{figure} 
\caption{Experimental scheme of the TXM.}
\includegraphics[scale = 0.32]{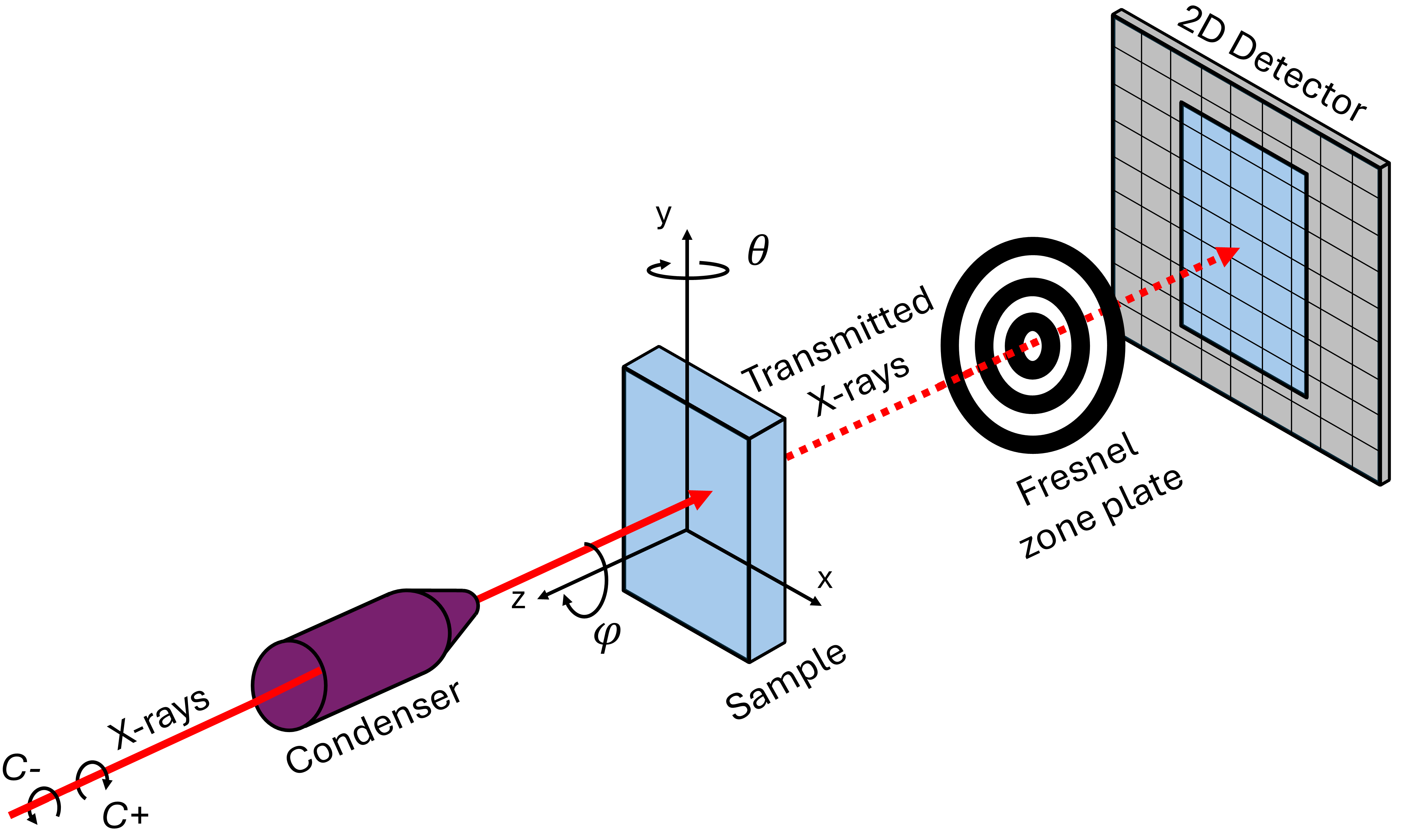}
\label{fig:microscope}
\end{figure}

The TXM allows imaging magnetic materials with a spatial resolution down to 30 nm depending on the Fresnel zone plate used and element sensitivity \cite{fischer2001resolutionTXM}. As depicted in Fig. \ref{fig:microscope}, the sample is illuminated either with $C+$ or $C-$ polarized X-rays coming from the condenser. The Fresnel zone plate lens collects the transmitted photons to form an image on a 2D detector. The incidence angle can be modified by rotating the sample around the vertical axis to a specific tilt angle ($\theta$). Additionally, it is possible to rotate the sample around its normal at $\theta = 0^\circ$ by $\varphi=90^\circ$.

During tomography, multiple repetitions of images or projections (P) are acquired to maximize the signal-to-noise ratio. For normalization, it is mandatory to measure images of the incoming beam without the sample, the so-called flat field (FF), analogous to the incident intensity in 2D. For samples with high absorbance, background images (BG) are also recorded. These BG images are measured by blocking the incident light with a specific highly absorbent sample and are employed to compensate for any possible stray light reaching the detector that could overcome the signal from very absorptive areas of the sample. Then, the transmittance ($T$) is calculated as follows:
  \begin{equation}
     T = \frac{P - BG}{FF - BG}
     \label{eq:normalization}
  \end{equation}
where all the used images (P, FF, BG) have been normalized to the electron current value in the storage ring, the exposure time, and the dark noise of the detector \cite{oton2015normalization}.
Transmittance projections are aligned to correct for drifts during acquisition and averaged to generate a single image for the same $\theta$, $\varphi$ and polarization.

The value of $T$ through a magnetic sample can be related to the sample properties by the Beer-Lambert equation taking advantage of the XMCD \cite{stohr2006magnetism}:
\begin{equation}
    T =\exp{\left(-\int{L\left(t\right)^{-1}\left(1+\delta\cdot\vec{k}\left(t\right)\cdot\vec{m}\left(t\right)\right)dt}\right)}
    \label{eq:Beer-Lambert}
\end{equation}
where $L(t)$ is the attenuation length of the X-rays in the sample, $\delta$ is the dichroic coefficient, and $\vec{k}\cdot\vec{m}$ is the dot product between the X-ray wave-vector and the magnetization. The transmittance is thus sensitive to the magnetization component parallel or anti-parallel to the X-ray beam direction. To recover the magnetization, projections are collected at different tilt angles ($\theta$ in Fig. \ref{fig:microscope}) using both circular polarizations. 

The magnetic signal can be obtained by operating with the transmittance for each polarization ($C+$ and $C-$) through the XMCD mechanism. Absorbance and XMCD images for each tilt angle $\theta$ are computed as:
\begin{equation}
    \text{Absorbance} = -\frac{1}{2} \ln{\left(T_{C+} \cdot  T_{C-}\right)} = - \int{L\left(t\right)^{-1}dt}
    \label{eq:absorbance}
\end{equation}
\begin{equation}
    \text{XMCD} = -\frac{1}{2} \ln{\left(\frac{T_{C+}}{T_{C-}}\right)} = -\int{L\left(t\right)^{-1}\left(\vec{k}\left(t\right)\cdot\vec{m}\left(t\right)\right)dt}
    \label{eq:xmcd}
\end{equation}
The absorbance provides information about the morphology of the sample while the XMCD is proportional to the magnetization. Before computing them, transmittance images of polarization $C+$ must be correctly aligned with their analogues of polarization $C-$ to avoid artifacts in the XMCD signal and, hence, in the 3D reconstruction of the magnetization.

The absorbance and XMCD images are both introduced as an input to the magnetic reconstruction algorithm. The algorithm chosen for this task depends on the dimensions of the sample. For quasi 2D materials, where its thickness is less than the axial resolution of the microscope ($\sim$60 nm), the algorithm detailed in \cite{herguedas2023scirep} can be used. It takes advantage of the small sample thickness to reduce the necessary number of angular projections (e.g. 6) and exposure time/number of images per angle and polarization. This decreases the acquisition time by a factor of 10 to 100 compared to standard 3D magnetic samples.

For thicker 3D systems,  to achieve good resolution, many more angular projections are required, as the in-depth resolution is directly related to the number of projections and the maximum tilt angle \cite{kak2001} (e.g. angular projections ranging from -55$^\circ$ to 55$^\circ$ in 87 steps \cite{hierro2020thinfilms}) and the images have to be tilt aligned to a common rotation axis before using the reconstruction algorithm. Iterative reconstruction algorithms based on X-ray tracing with gradient descent optimization \cite{donnelly2018newjurphy, pam2024resire} or algebraic methods \cite{hierro2018journalsynchrotron} can be used to recover the 3D magnetic configuration. In these algorithms, a guess object is created, which for the initialization could be a 3D matrix of zeros or a random magnetization. For each iteration, calculated projections from the guess object are compared to the measured ones in order to obtain an error, which is then used to update the guess object.

Furthermore, due to the XMCD contrast mechanism, a single TS allows recovering two magnetization components, the out-of-plane ($m_z$ in the frame of reference of the sample in Fig. \ref{fig:microscope}) and one of the in-plane ($m_x$ in Fig. \ref{fig:microscope}). The other in-plane component ($m_y$) is obtained by rotating the sample around $\varphi$, as shown in Fig. \ref{fig:microscope}, providing an orthogonal TS. 
Once the absorbance and magnetic reconstructions are obtained for both TS (TS1 with $\varphi = 0^\circ$ and TS2 with $\varphi = 90^\circ$), they are merged to get the three components of the magnetization vector. From the 3D reconstructed absorbance, a 3D mask is created for each TS to obtain the geometrical transformation between both to a common coordinate system. This transformation is then applied to $m_y$, which is reconstructed from TS2. Finally, the three components of the magnetization $m_x$, $m_y$, and $m_z$ are obtained.

\section{Software structure}
MARTApp consists of four main data treatment steps developed using Python \cite{python} and two iterative reconstruction algorithms implemented using MATLAB \cite{matlab,hierro2018journalsynchrotron,herguedas2023scirep} but compiled to be distributed together with the application. The GUI has been built using the PyQt5 library \cite{pyqt5} and Qt designer \cite{qt}. For its distribution, it comes in two flavours within two different docker images: a native one that uses a Debian-Linux base image as the operative system and needs package X11, and another one that includes noVNC \cite{novnc} to work directly in the browser without needing graphic interface forwarding. Both docker images allow the user to be able to use Windows, Mac, or Linux.

\begin{figure} 
\centering
\caption{Main screen of MARTApp.}
\includegraphics[scale = 0.6]{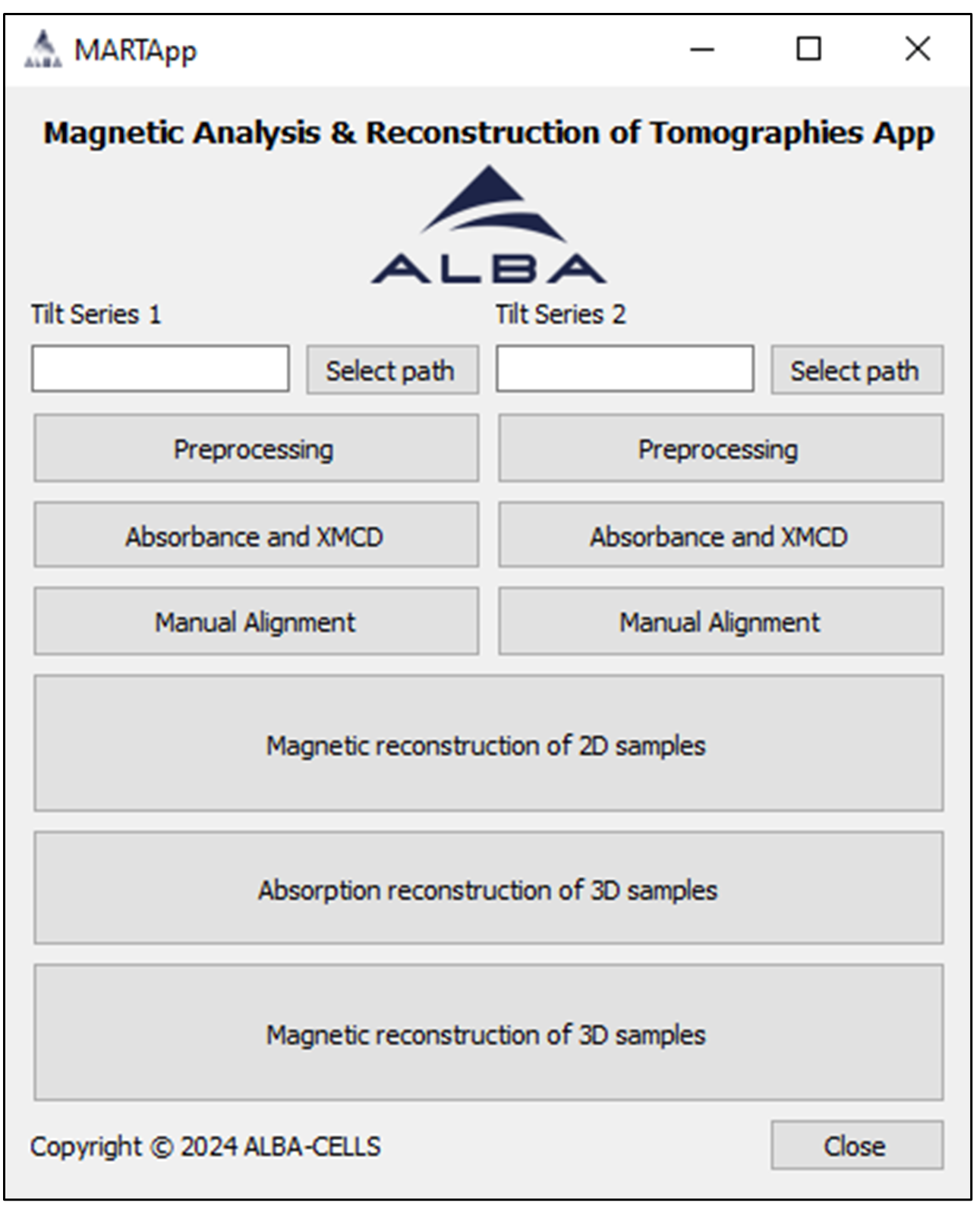}
\label{fig:main_MARTApp}
\end{figure}

The GUI presents a main window or widget (see Fig.\ref{fig:main_MARTApp}) where the two required TS may be processed in parallel, allocating each one to a different directory. 

Once both TS are selected, we can start the processing pipeline which has consecutive steps ending with the absorbance and magnetic 3D reconstructions. MARTApp follows the workflow described above for MVT and includes the following modules: 1) pre-processing (data reading and transmittance computation from single raw images), 2) absorbance and XMCD signal (image alignment for each polarization, absorbance and XMCD computation), and 3) magnetic reconstruction of quasi-2D or 3D samples. Additionally, it also includes a module for manually correcting or improving the automatic 2D alignment of the images. Finally, every module allows the user to inspect the intermediate and final results using Matplotlib \cite{hunter2007matplotlib} or Napari \cite{chiu2022napari}. Moreover, the different step modules of the GUI can be used independently.

The analysis workflow may start with the file format of the MISTRAL-TXM software (e.i. XRM files from Zeiss-Xradia) using the pre-processing module or with the normalized transmission images in HDF5 format (automatic conversion from XRM to HDF5 available) using the absorbance and XMCD module.

In the following, the modules are described in detail.

\subsection{Pre-processing}
\begin{figure} 
\caption{Interface and workflow of the pre-processing module.}
\includegraphics[scale = 0.5]{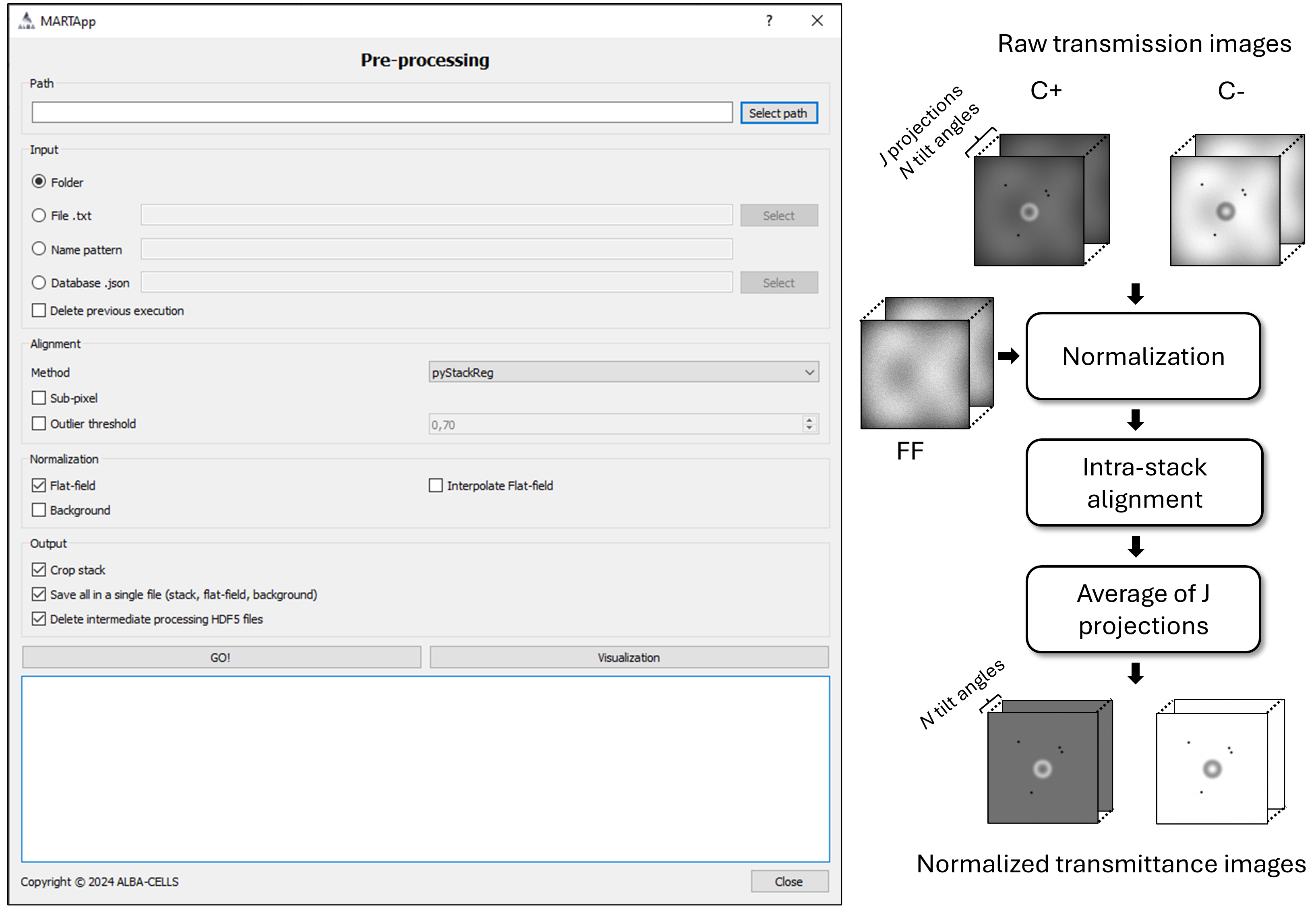}
\label{fig:gui_preprocessing}
\end{figure}

  The pre-processing module allows reading the raw images of a TS obtained from the microscope and outputs the transmittance images. Fig. \ref{fig:gui_preprocessing} shows the interface and the workflow of the module. The raw transmission images and metadata are read from the XRM files and the different projections are grouped by tilt angle and polarization in two corresponding stacks $C+$ and $C-$. Then, each group of 
 $J$ projections is normalized using Eq.\ref{eq:normalization}.

  For those cases where FF have not been acquired for each angular projection, this module includes an option that allows interpolating the measured FFs to estimate the missing ones. Note that for long acquisition times of several hours, the incident flux and its spatial distribution may vary and an accurate normalization is needed.
  
  For the 2D alignment step of the $J$ normalized projections, the module includes four different alignment algorithms allowing for pixel-wise or subpixel registration:
  \begin{itemize}
      \item \textbf{Cross-correlation in real space}: Template matching-based algorithm using OpenCV and cross-correlation in real space \cite{opencv2012intro,opencv2006crosscorr}.
      \item \textbf{Cross-correlation in Fourier}: Scikit-image implementation of the cross-correlation alignment in Fourier space \cite{guizar2008crosscorrfourier, van2014scikit}.
      \item \textbf{Optical flow in Fourier}: Optical flow-based alignment algorithm in Fourier space \cite{of1995}.
      \item \textbf{PyStackReg}: Python/C++ port of the ImageJ extension TurboReg/StackReg \cite{thevenaz1998stackreg}, a pyramid subpixel registration based on intensity, widely used in time-resolved fluorescence and wide-field microscopy.
  \end{itemize}
 After the 2D alignment, angular projections are averaged, resulting in one transmittance image per each tilt angle and polarization. The results are saved in two different HDF5 files, one per polarization $C+$ and $C-$, containing the transmittance TS, the FF and BG, the angles, and other metadata.

\subsection{Absorbance and XMCD signals}
\begin{figure} 
    \caption{Interface and workflow of the absorbance and XMCD signals module.}
    \includegraphics[scale = 0.4]{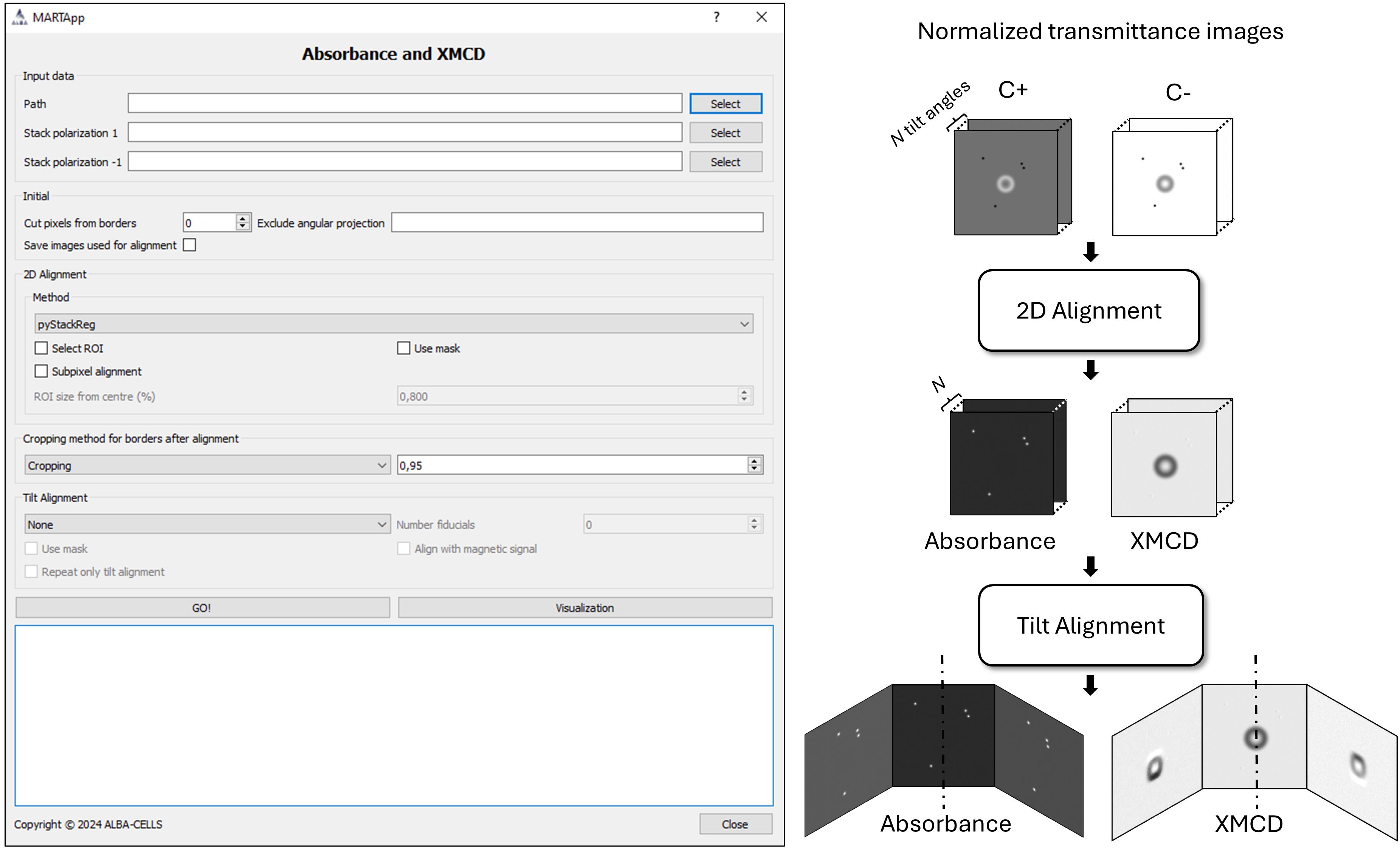}
    \label{fig:gui_xmcd}
    \end{figure}

    The second module, shown in Fig. \ref{fig:gui_xmcd}, allows the computation of the absorbance and XMCD images from the normalized images of both polarizations contained in the HDF5 files. Firstly, it allows to crop pixels from the border of the images to remove empty pixels generated by the precedent 2D alignment procedure. Moreover, it is possible to exclude angular projections from the workflow that might have poor quality due to punctual vibrations, injections, shadow effects due to rotation, etc. 

Then, angular projection series with the same tilt angle from $C-$ polarization are aligned with their corresponding $C+$ ones. Here, in addition to the above-mentioned 2D alignment algorithms, the alignment includes an option to select a region of interest (ROI) and/or the creation of a binary mask to suppress the magnetic signal of the images which could, otherwise, affect the alignment process due to the opposite contrast of $C+$ and $C-$ polarizations. Again, images can be cropped or filled with a certain value after the 2D alignment in order to remove shift artifacts. 
Moreover, since the polarization fraction between $C+$ and $C-$ reaching the sample might not be strictly equivalent, an intensity correction factor can be applied between opposite polarization images by selecting a ROI where no magnetic contrast is present. This factor is then applied to the $C-$ polarization images to ensure that the average intensity within the field of view is the same for both polarizations. Then, the absorbance and XMCD stacks are computed by Eq. \ref{eq:absorbance} and Eq. \ref{eq:xmcd} respectively. 
    
Before the 3D magnetic reconstruction, it is mandatory to align the images to a common rotation axis and MARTApp includes three algorithms for this task:
\begin{itemize}
    \item \textbf{CT align}: Fiducial-based tilt alignment pipeline using IMOD's tilt alignment algorithm \cite{kremer1996imod, mastronarde1997imod}.
    \item \textbf{Optical flow in Fourier}: Same algorithm as for 2D alignment but performing a subsequent alignment, i.e. each angle is aligned to the previous one.
    \item \textbf{PyStackReg}: Same algorithm as for 2D alignment but enabling the tilt alignment option.
\end{itemize}
It is important to note that the Optical flow in Fourier and PyStackReg algorithms rely on nominal tilt angles due to the absence of computed tilt angles.
Usually, the alignment is performed on the absorbance stack and the same shifts are then applied to the XMCD one. To facilitate this step, more options have been included, such as the creation of a mask from the absorbance series to isolate the sample from the rest of the field of view or the use of the XMCD stack when the magnetization is mainly out-of-plane and the contrast does not change when rotating. The output HDF5 file contains the original and the 2D-aligned transmittance images for each polarization, the XMCD and absorbance signal, and, if selected, the tilt-aligned datasets. 

\subsection{Manual image alignment}
    Due to the different type of samples and their structural complexity, automatic 2D algorithms may not perfectly align the images with the needed accuracy to compute the absorbance and XMCD signal stacks. In order to solve this problem, MARTApp includes a module to manually align the two polarization stacks (or any set of images in a HDF5 file). In this module, the user selects the reference and it applies translations, rotations, scaling, and intensity factors to the other image. Both the overlap and the difference of the two images are shown to evaluate the applied alignment. Once the user checks the alignment is correct, the transformed image or stack is exported into a HDF5 file with their corresponding absorbance and XMCD signals. Afterwards, the tilt alignment can be performed returning to the XMCD module.

\subsection{Magnetic reconstruction of quasi-2D samples} 
    This module implements the reconstruction algorithm described in \cite{herguedas2023scirep} for quasi-2D samples or thin films (thickness $\leq60$ nm). After obtaining the tilt-aligned absorbance and XMCD stacks, MARTApp applies affine transformations to the images to correct the cosine stretching. The computation of the transformation is done by registering the absorbance of the angular projections with the one at normal incidence. An initial transformation can be obtained by selecting points manually on the absorbance image at normal incidence and using the nominal tilt angle. Then, both TS are merged and the rotation angle $\varphi$ is computed using again the absorbance images of each TS at normal incidence. The same transformations are applied to the XMCD stack. In the case of quasi-2D multilayers, an additional translation of the magnetic signal along the sample depth, proportional to the tilt angle and the distance between the absorbance layer and the magnetic one, can be performed. This is included for a correct alignment in a continuous film, where it is mandatory to have landmarks for the alignment. Finally, the magnetic signal is introduced in the reconstruction algorithm and the output is a HDF5 file containing the 3D magnetization vector of the quasi-2D system under study.

\subsection{Absorbance and magnetic reconstruction of 3D samples}
\begin{figure} 
    \caption{Interface and workflow of the absorbance reconstruction module for 3D magnetic materials.}
    \includegraphics[scale = 0.45]{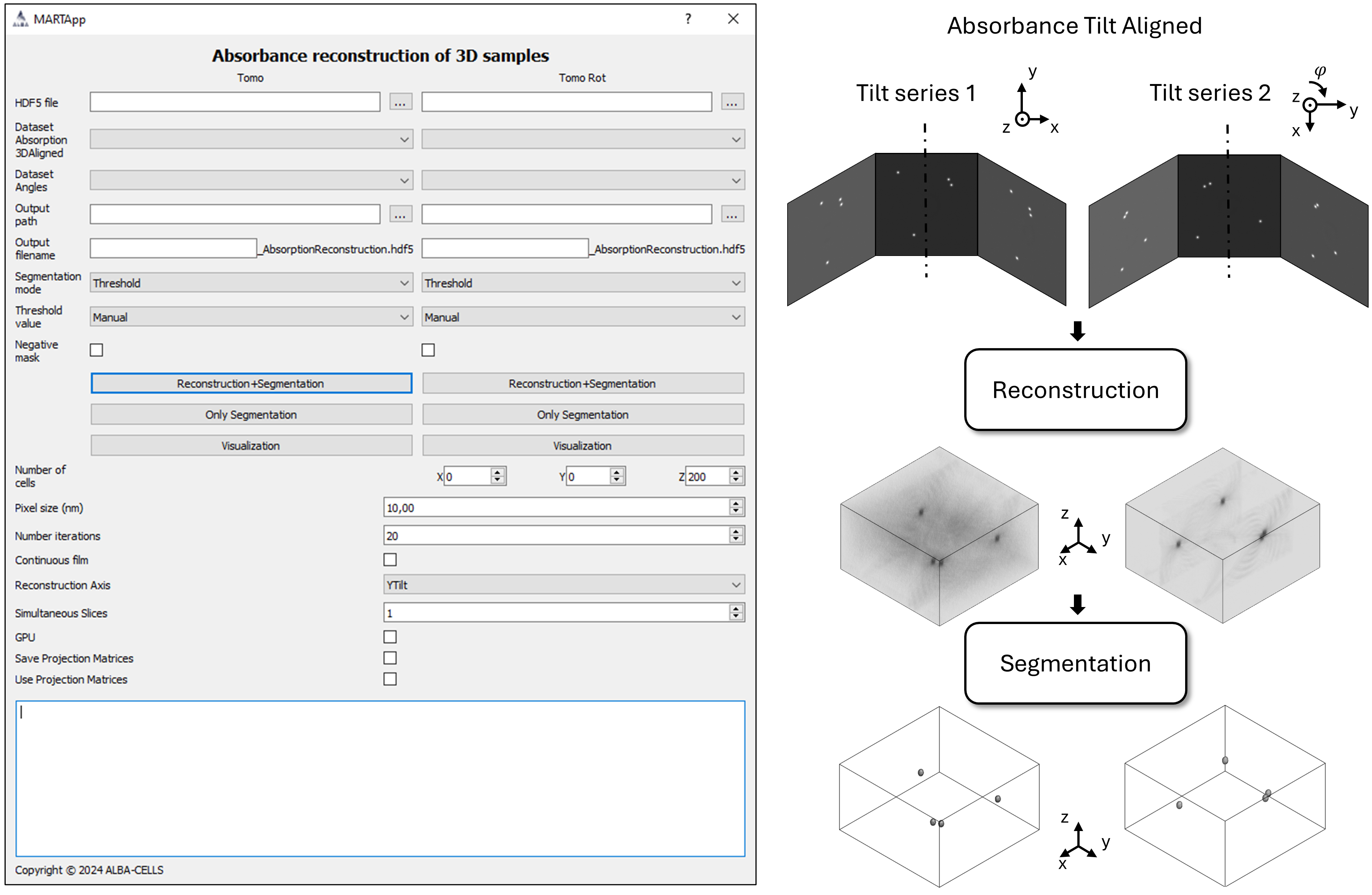}
    \label{fig:gui_absorbance3D}
    \end{figure}

    This last module is divided into two parts, absorbance and magnetic reconstruction of 3D samples and uses the algorithm described in \cite{hierro2018journalsynchrotron}.
    The interface and scheme of the first part can be found in Fig. \ref{fig:gui_absorbance3D}. First, the 3D absorbance reconstruction is obtained from the corresponding tilt-aligned images of each TS. This allows us to obtain the sample morphology, i.e. the 3D map of the non-magnetic term of the attenuation coefficient in Eq. \ref{eq:Beer-Lambert}. To run this reconstruction algorithm, several input parameters are required: the size of the reconstructed volume, the pixel size of the images, the number of iterations, the rotation axis orientation, and whether the sample is a continuous film or not. The continuous film flag extends the reconstruction model horizontally (vertically) for TS where the rotation axis is around the $Y$ ($X$) axis (see Fig. \ref{fig:microscope}), simulating an infinite film to minimize edge artifacts. After the absorbance reconstruction, a segmentation is performed to create a 3D mask which will be used later to find the geometrical transformation between TS1 and TS2. Three options related to the type of sample have been included for segmentation: a threshold for general systems, the application of a 2D mask for nanostructures with 2D symmetry, and the creation of a mask for continuous films from Au nanoparticles deposited on top of the magnetic film for alignment purposes \cite{hierro2020thinfilms}. For each one, the mask can be generated using different automatic thresholding methods: Otsu \cite{otsu1975threshold}, Huang \cite{huang1995thresholding}, MaxEntropy \cite{kapur1985maxentropy}, or Triangle \cite{zack1977triangle}; or by manual selection in a reconstructed signal intensity histogram. The output of each TS segmentation is two binary masks, one to be used as a constraint in the magnetic reconstruction (optional), and another one to find the geometric transformation between the reconstructed volumes of both TS to obtain the three magnetization components on the same reference frame.
    
\begin{figure} 
    \caption{Interface and workflow of the magnetic reconstruction module for 3D materials.}
    \includegraphics[scale = 0.4]{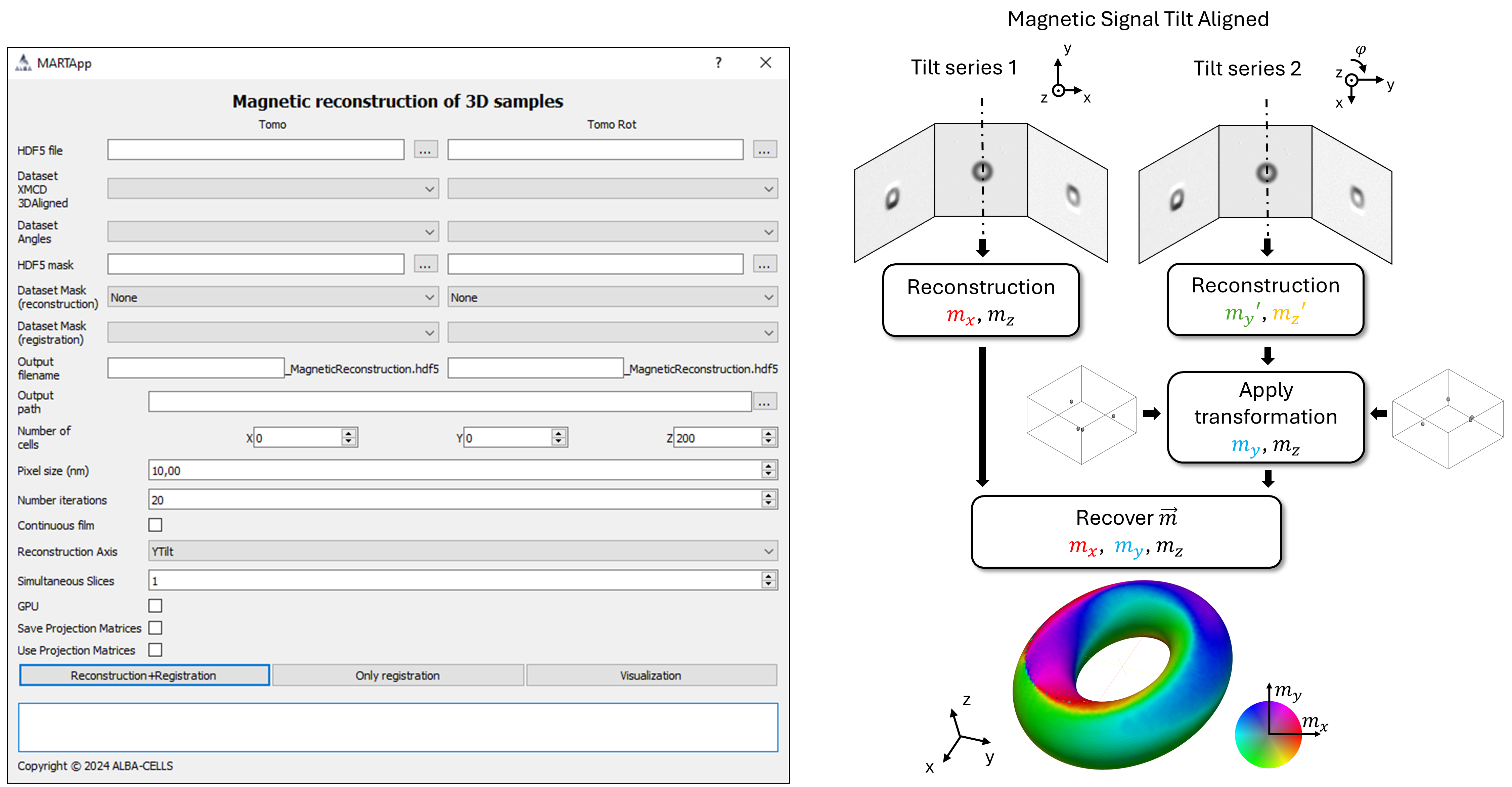}
    \label{fig:gui_mag3D}
    \end{figure}

    In the second part, shown in Fig. \ref{fig:gui_mag3D}, the magnetization for each TS is reconstructed from the XMCD tilt-aligned stack. The same parameters mentioned above (size of the reconstructed volume, pixel size, etc.) must be introduced in the algorithm. The magnetic reconstruction of TS1 outputs $m_x$ and $m_z$, whilst TS2 reconstruction outputs $m_y'$ and $m_z'$. Then the geometrical transformation between both TS is computed with the binary 3D mask obtained from the previous step as already mentioned. An initial estimation of the transformation can be obtained by selecting three points in the reconstructed absorbance space or introducing a rotation angle around the X-ray direction of propagation (see Fig. \ref{fig:microscope}). Finally, the transformation is then applied to $m_y'$ to obtain $m_y$ in the same reference system. For consistency between the components reconstructed from TS1 and TS2, $m_z$ is used as a normalization parameter as it must be equal for both TS. After applying this correction, the 3D magnetization vector is obtained, where $m_x$ and $m_z$ are reconstructed from TS1 and $m_y$ is computed by rotating TS2. 
    
\section{Use cases}
MARTApp can be used and has been tested at the MISTRAL beamline at ALBA with different kinds of samples, e.g. quasi-2D systems, thin and continuous films or nanostructures. We foresee that any possible magnetic sample system that can be described by the Beer-Lambert equation (Eq. \ref{eq:Beer-Lambert}) can be analyzed with MARTApp. The set of 2D and rotation alignment algorithms, helping features (e.g. intensity correction between polarization), and reconstruction algorithms provide adaptability and make MARTApp highly beneficial for the user community.

With the aim of demonstrating MARTApp capabilities, we present in what follows a practical example.

\subsection{Simulation of a magnetic vectorial tomography}
\begin{figure} 
\caption{Scheme of the simulation volume. The hopfion is contained in the cylinder. In the rest of the space, the magnetization is along the Z direction, shown in red arrows. Black spheres represent the Au fiducial markers. Right panels show the top and side views.}
\includegraphics[scale = 0.3]{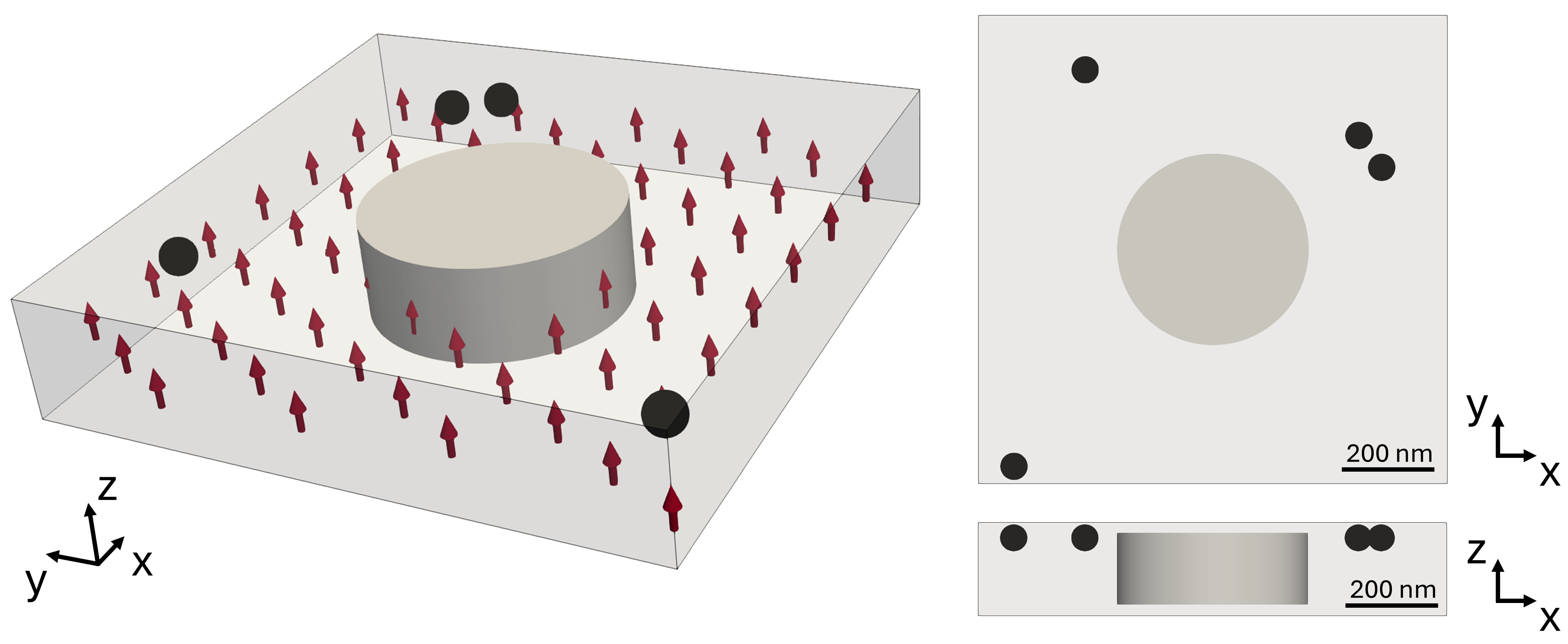}
\label{fig:hopfion_description}
\end{figure}

Herein, we have simulated a magnetic hopfion within a continuous film as a model system. The hopfion can be described as a generalization of a skyrmion in the 3D space where the magnetization forms twisted loops around the centre \cite{rybakov2022hopfdescription,voinescu2020hopfdescription}. Although the hopfion is of significant interest to the nanomagnetism and spintronics community due to its unique properties and potential applications in data storage devices \cite{wang2019hopf}, it is chosen here primarily for its complex 3D character. This allows us to demonstrate the capabilities of MARTApp in recovering arbitrary three-dimensional magnetization configurations. 
The analytical expression used for the simulation was extracted from \cite{sutcliffe2018hopfion}. The hopfion has been computed in a cylinder with a height of 170 nm and a diameter 510 nm surrounded by homogeneous out-of-plane magnetization within a continuous film of 200 nm thick. Spheres that act as Au fiducial markers of 100 nm diameter have been introduced in the simulated space. A scheme of the cylinder containing the hopfion and the fiducials in the simulation space can be found in Fig. \ref{fig:hopfion_description}.
  
From the simulated magnetization, we have obtained the projections by calculating the trajectory of the X-ray beam through the space and applied the Beer-Lambert equation (Eq. \ref{eq:Beer-Lambert}) to calculate the intensity at each pixel of the detector. The number of cells is 1200x1200x250 with a voxel size of 1 nm and the parameter $L$ was set to $4$ $\mu$m$^{-1}$, similar to the Fe, and $\delta = 1$.
Projections have been computed for different tilt angles, ranging from $-65^\circ$ to $65^\circ$ in steps of $2^\circ$ to simulate the constraint of having a missing wedge in a continuous film system. Two orthogonal TS have been simulated for which the sample has been rotated $\varphi = 90^\circ$.

Then, to mimic the experimental transmission images that can be obtained from the microscope, each projection has been convolved with a Gaussian function (e.g. standard deviation of 5)  to take into account the dispersion caused by the lens. The projections were then resized to a match the pixel size of the experimental data (e.g. 10 nm). The angular projections were obtained by averaging 20 projections with Poisson noise. Finally, random shifts have been added to generate misalignment and drift effects between different tilt angles.

\subsection{Processing and reconstruction of the hopfion}
    In the following, we describe the processing, analysis, and reconstruction of the simulated hopfion tomography. The workflows of figures \ref{fig:gui_xmcd}, \ref{fig:gui_absorbance3D} and \ref{fig:gui_mag3D} display the process and output images or volumes obtained at each step.

    Transmittance images for polarization stacks $C+$ and $C-$ are introduced in the computation of the XMCD and absorbance module. For both TS this step is the same: 20 pixels were cropped from the borders and the algorithm chosen to perform the 2D alignment is sub-pixel Cross-correlation in Fourier space, with a ROI size from the centre of 80\% of the field of view. The raised cosine radial mask specific to this algorithm is set to default values to avoid artifacts. No intensity correction factor has been applied between polarizations, as no difference between the polarization intensity $C+$ and $C-$ has been introduced in the simulated images. After the alignment, the size of the image is trimmed by 25\% to avoid edge displacements. The absorbance and XMCD signals are computed and used for the tilt alignment, performed with CT align.

    Once the absorbance and XMCD stacks are tilt aligned, the 3D reconstructions (absorbance and magnetic) are performed in a space of 440x440x250 voxels of 10 nm size with the option for reconstructing continuous film activated. The rotation is performed around the $Y$ axis, i.e. the vertical direction of the image (see Fig. \ref{fig:hopfion_description}). The reconstruction algorithm runs for 50 iterations. Once the 3D absorbance reconstruction is obtained, it is manually (histogram) thresholded to obtain a mask of the fiducial markers for each TS. With these masks, the transformation between TS1 and TS2 reconstructed volumes is computed. Finally, the magnetic reconstruction is performed with the same parameters as the absorbance one and the GUI outputs the magnetization vector.

\begin{figure} 
    \caption{a) Ground truth. b) 3D reconstructed hopfion from the simulated transmission images of the ground truth. The first row shows the in-plane orientation of the magnetization in the middle layer of the XY plane. Each row represents the magnetization $m_x$, $m_y$, and $m_z$, for the XZ plane respectively.}
    \includegraphics[scale = 0.35]{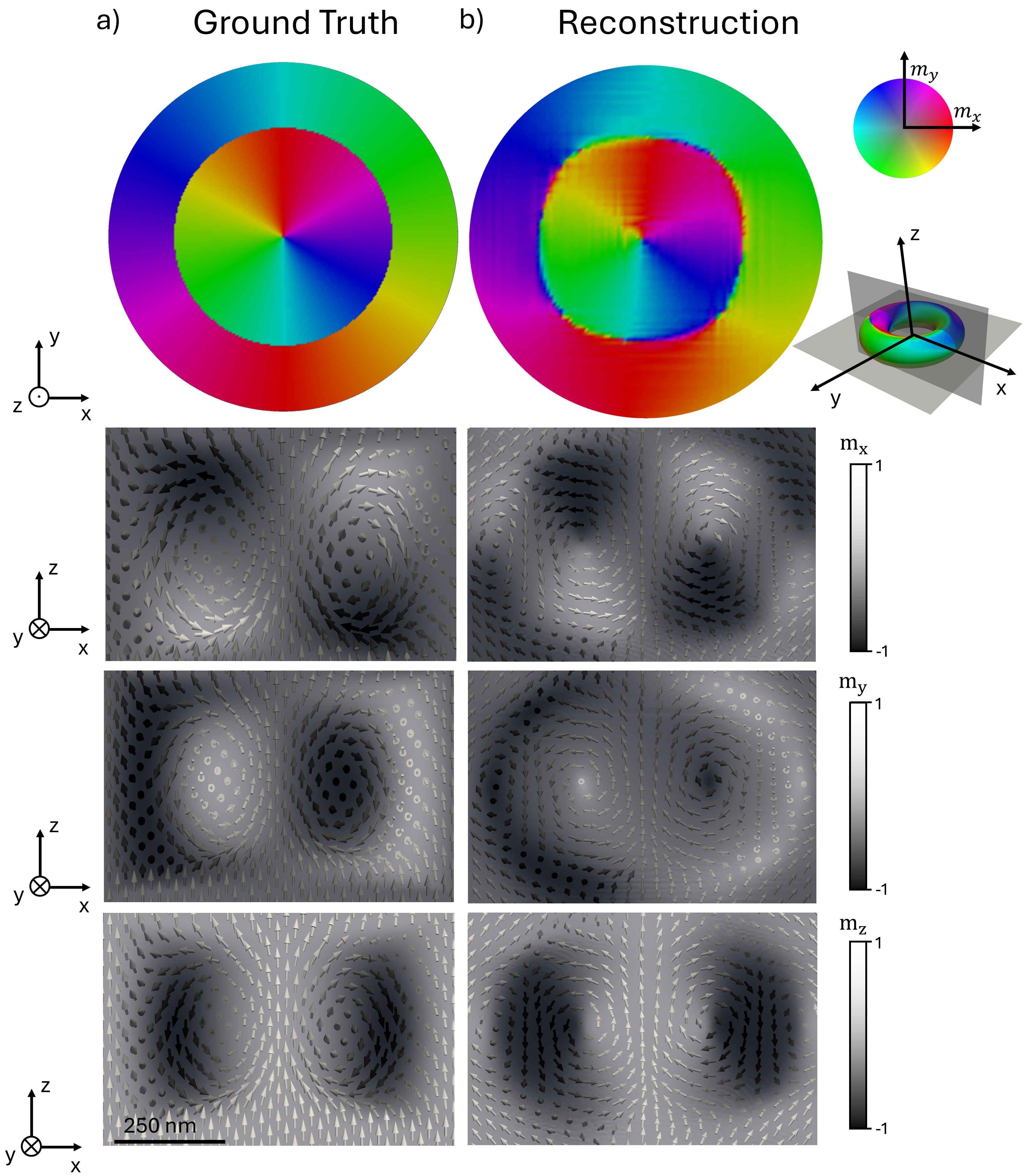}
    \label{fig:hopfion_GT}
    \end{figure}

    The results of the reconstructed magnetization of the hopfion can be found at the end of the workflow in Fig. \ref{fig:gui_mag3D}. It has a toroidal shape where the magnetization at its centre is oriented along the Z direction. Around this toroidal region, the magnetization is twisting.
    Fig. \ref{fig:hopfion_GT} shows the comparison between (a) the ground truth and (b) the reconstruction. The first row of the images presents the in-plane components of the magnetization of the XY plane, showing the rotation of the magnetization. The other three rows represent the three components of the magnetization, $m_x$, $m_y$, and $m_z$ for the XZ plane respectively. The reconstruction of the hopfion in this plane is affected by the missing wedge configuration, leading to the introduction of a magnetic signal at the borders of the hopfion. However, the characteristics of the magnetization texture are clearly recovered with the use of MARTApp.

\section{Conclusions and future work}
Herein we have presented MARTApp, a cross-platform application that allows processing, analyzing, and reconstructing the magnetic tomographic data from transmission X-ray microscopy experiments. The four different available modules to process magnetic tomography have been described. They involve normalization, computation of the absorbance and XMCD signals, and reconstruction of the magnetization. They have been designed to work with different kinds of samples, from quasi-2D systems to 3D nanostructures including continuous films. Finally, a synthetic system consisting of a hopfion in a continuous film has been used to demonstrate the utilization of the application. This tool is well suited to help expert and non-expert users in the analysis of MVT and it provides an easy and fast way to obtain 3D magnetization vector reconstructions.

In the future, MARTApp will be adapted to work with other data standards such as NeXus data format \cite{konnecke2015nexus}. In addition, more features specific to other types of user samples will be continously added as they are being analyzed and new requirements will emerge. MARTApp is highly flexible and can be adapted easily making it a valuable tool to perform MVT for the magnetism community.

\section{Software and data availability}

MARTApp is available to download, accompanied by manuals and instructions at the GitHub repository \url{https://github.com/ALBA-Synchrotron/MARTApp}. It is distributed in its two flavours (native and web-based GUIs) through two Docker images that can be pulled from the GitHub registry. The data used to demonstrate the software functionality for this work is open source and is available in Zenodo \cite{herguedas_2024_14094130}.

     % Appendices appear after the main body of the text. They are prefixed by
     % a single \appendix declaration, and are then structured just like the
     % body text.

% \appendix
% \section{Appendix title}

% Text text text text text text text text text text text text text text
% text text text text text text text.

% \subsection{Title}

% Text text text text text text text text text text text text text text
% text text text text text text text.

% \subsubsection{Title}

% Text text text text text text text text text text text text text text
% text text text text text text text.

     %-------------------------------------------------------------------------
     % The back matter of the paper - acknowledgements and references
     %-------------------------------------------------------------------------

     % Acknowledgements come after the appendices

\ack{Acknowledgements}

ALBA Synchrotron Light Source is funded by the Ministry of Research and Innovation of Spain, by the Generalitat de Catalunya, and by European FEDER funds. This work has been supported by Spanish MICINN under grant PID2022-136784NB/ AEI/10.13039/501100011033, and by Asturias FICYT under grant AYUD/2021/51185 with the support of FEDER funds. A. E. H.-A. acknowledges the support from the Severo Ochoa Predoctoral Fellowship Program (nos.PA-23-BP22-093) from the Government of the Principality of Asturias (Spain). The authors would like to thank the ALBA IT section for the support provided in hardware infrastructure and all those involved in the operation of ALBA.

     % References are at the end of the document, between \begin{references}
     % and \end{references} tags. Each reference is in a \reference entry.

% \begin{references}
% \end{references}
%\bibliographystyle{plain} % We choose the "plain" reference style
\bibliographystyle{IEEEtran}
\bibliography{iucr} % Entries are in the refs.bib file

     %-------------------------------------------------------------------------
     % TABLES AND FIGURES SHOULD BE INSERTED AFTER THE MAIN BODY OF THE TEXT
     %-------------------------------------------------------------------------

     % Simple tables should use the tabular environment according to this
     % model

% \begin{table}
% \caption{Caption to table}
% \begin{tabular}{llcr}      % Alignment for each cell: l=left, c=center, r=right
%  HEADING    & FOR        & EACH       & COLUMN     \\
% \hline
%  entry      & entry      & entry      & entry      \\
%  entry      & entry      & entry      & entry      \\
%  entry      & entry      & entry      & entry      \\
% \end{tabular}
% \end{table}

     % Postscript figures can be included with multiple figure blocks

\end{document}